\documentclass[12pt]{article}
\usepackage{amssymb}
\usepackage{amsmath}
\usepackage{eucal}

\def\bea{\begin{eqnarray}}
\def\eea{\end{eqnarray}}
\def\be{\begin{equation}}
\def\ee{\end{equation}}
\def\re#1{(\ref{#1})}
\def\bx{\bar{x}}

\begin{document}

\begin{titlepage}
\rightline{file: vectorfinal.tex}
\rightline{PACS numbers: 04.50.Cd}
\vskip 3. cm
\begin{center}
\huge{\bf{Kaluza-Klein towers for real vector fields in flat space
}}
\end{center}
\vskip 0.6 cm
\begin{center} {\Large{Fernand Grard$^1$}},
\ \ {\Large{Jean Nuyts$^2$}}
\end{center}
\vskip 1 cm

\noindent{\bf Abstract}
\vskip 0.2 cm
{\small
\noindent
We consider a free real vector field propagating in a five dimensional flat space
with its fifth dimension compactified either on a strip or on a circle and
perform a Kalaza Klein reduction which breaks SO(4,1) invariance while preserving 
SO(3,1) invariance. 
Taking into account the Lorenz gauge condition,
we obtain from
the most general hermiticity conditions
for the relevant operators
all the allowed boundary conditions which have to be imposed on the
fields in the extra-dimension.
The physical Kaluza-Klein mass towers, which result in a four-dimensional brane,
are determined in the different distinct allowed cases. 
They depend on the bulk mass, on the parameters of the boundary conditions 
and on the 
extra parameter present in the Lagrangian. In general, they involve vector states
together with
accompanying scalar states.  
 }
\vfill
\noindent
{\it $^1$  fernand.grard@umons.ac.be,  Physique G\'en\'erale et
Physique des Particules El\'ementaires,
Universit\'e de Mons - UMONS, 20 Place du Parc, 7000 Mons, Belgium}
\vskip 0.2 cm
\noindent
{\it $^2$  jean.nuyts@umons.ac.be,
Physique Th\'eorique et Math\'ematique,
Universit\'e de Mons - UMONS,
20 Place du Parc, 7000 Mons, Belgium}
\end{titlepage}

\newpage
\section{Introduction \label{Intro}}

In previous publications \cite{GN1}, we have applied the Kaluza-Klein reduction
procedure \cite{KK} to various fields propagating in a five dimensional space
with the fifth dimension compactified either on a strip or on a circle. 
 
Our approach is based on an exhaustive study of the hermiticity properties
of the operators defining the masses. This results in the determination of all the 
sets of allowed field boundary conditions which have to be applied in the extra dimension. 
We have successively considered scalar and spinor fields in flat and warped spaces
and have established in each case all the corresponding Kaluza-Klein towers. 
Our main result is that, apart from the usual towers where the masses (or masses squared) 
are regularly spaced (essentially linearly), 
there appears new types of towers with masses which are solutions 
of specific transcendental functions.  

We consider here the case of a free real vector field in the five 
dimensional flat space and discuss its most general free Lagrangian
and related equations of motions (see Sect.\re{massiveeq}). 
Requiring the hermiticity of the relevant operators,
the basic boundary relations for the field are established
in Sect.\re{flatBC}. 
Adopting
a K-K reduction which breaks SO(4,1) covariance while preserving
SO(3,1) covariance, the boundary relations for the reduced fields in the 
fifth dimension are derived
(see Sect.\re{flatKKreduc}), from which, taking into account the 
restrictions imposed by the 5-dimensional Lorenz gauge condition,
all the sets of allowed boundary conditions are obtained(see \re{flatKKeq}). 
Some physical considerations 
about the locality of the boundary conditions 
and
about the possible extension to a closed circle in the extra-dimension
are given in Sect.\re{locality} while the physical 
relevance of the 
mass scales which have to be adopted in a flat space 
is discussed in Sect.\re{physmass}.
A complete analysis of the boundary conditions
leads to the
appearance of both 4-dimensional vector and scalar states 
in the
resulting Kaluza-Klein mass towers 
as seen in the 4-dimensional branes(see Sect.\re{towers}).

Some physical consequences
are discussed and the main results 
are summarized in the 
conclusions (Sect.\re{conclusion}) 


\section{Real vector field in five dimensional flat space. 
Equations of motion. 
Kaluza-Klein reduction. 
Boundary relations and boundary conditions
\label{flatmass}}

\subsection{Lagrangian. Equations of motion \label{massiveeq}}

Let $B^{A}$ be a free real vector field propagating in a
five-dimensional flat space
\be
x^{A}=\{\bx,x^5\}=\{x^{\mu},x^5\},\ \mu=0,1,2,3,\ s=x^{5}
\label{variables}
\ee 
with an infinite four dimensional Minkowski space $\bx$ 
and the fifth dimension $s$ compactified on a strip $[0, 2 \pi R]$ or on a
circle of radius $R$. The metric is
$\eta ^{\mu\nu}={\rm{diag}}(1,-1,-1,-1),\ \eta^{\mu 5}=0,\  \eta^{55}=-1$.
The most general Lorentz invariant action, quadratic in the 
vector field $B^{A}$, contains, apart from the bulk mass
$M$, two (real)
free parameters $\lambda_1$ and $\lambda_2$ 
\be
{\cal{A}}
 =\int \biggl(\frac{1}{4}F_{AB}F^{AB}
 +\frac{\lambda_1}{2}(\partial_AB^A)^2
 +\frac{\lambda_2}{2}(\partial_A B_B)(\partial^B B^A)
 -\frac{M^{2}}{2}B_{A}B^{A}\biggr) d^{5}x
\label{lag2}\ee
with
\be
F_{AB}=\partial_A B_B-\partial_B B_A.
\label{defF}
\ee
Since the two factors of $\lambda_1$ and $\lambda_2$ are equal up to a divergence, the equations of motion
\re{square5b} depend only on $\lambda$ with
\be
\lambda=\lambda_1+\lambda_2.
\label{lambdadef}
\ee 

The so-called usual case corresponds to the choice $\lambda=0$ and we will focuss our attention on it. 
We will also discuss what we call 
the generalized case with $\lambda\neq 0$. Compared to the usual case, the generalized case 
involves, as we will explain later, essentially
one more degree of freedom which is of scalar type. We will see 
that the five dimensional vector field leads, upon Kaluza-Klein reduction, to vector and scalar particles, mimicking, 
in some cases, 4-dimensional situations corresponding to a generalized case in the brane.

\vskip 3 cm
\subsubsection{Equations of motion. Usual case ($\lambda= 0$) \label{usualeqgen}}
\vskip 0.3 cm

\noindent The usual Lorentz invariant action 
(see however the discussion of the ``Generalized case'' below)
corresponds to the particular choice $\lambda=0$
\be
{\cal{A}}
 =\int \left(\frac{1}{4}F_{AB}F^{AB}
 -\frac{M^2}{2}B_{A}B^{A}\right) d^{5}x
\label{lag1}
\ee
and leads to the equations of motion 
\be
(\square_5+M^2)B^A-\partial^A(\partial_C B^C)=0.
\label{square5a}
\ee

\vskip 0.3 cm
\noindent{\bf{Usual Case ($\lambda=0$) with  $M^2\neq 0$}} 
\vskip 0.3 cm
\noindent For $M^2\neq 0$, one deduces 
from the divergence of \re{square5a}
the usual Lorenz condition
\begin{equation}
\partial _{A}B^{A}=0
\label{gauge5}
\end{equation}
and the equations of motion
\begin{equation}
\left(\square _{5}+M^2\right)B^{A}=0,
\label{square5}
\end{equation}
which justify the interpretation of $M^2$ as the bulk mass squared
of the transverse (Eq.\re{gauge5}) massive (Eq.\re{square5})
vector field with four degrees of freedom.

\vskip 0.3 cm
\noindent{\bf{Usual Case ($\lambda=0$) with $M^2= 0$}} 
\vskip 0.3 cm
\noindent 
For $M^2=0$, the Lagrangian is invariant 
under the gauge transformation
\be
B^{A}\longrightarrow B^{A}+\partial^{A}\Lambda
\label{gaugenorm}
\ee
for any arbitrary $\Lambda$.

The Lorenz condition
\re{gauge5}
can be obtained by a suitable gauge transformation 
\be
(B')^A=B^A+\partial^A\Lambda
\label{gaugetf}
\ee
leading to
\bea
\square_5 B^{'A}&=&0
   \label{square5m0}\\
\partial_A B^{'A}&=&0
   \label{gauge5m0}  
\eea
provided $\Lambda$ is chosen as
\be
\square_5\Lambda=-\partial_AB^A
\label{Lambdaeq}
\ee
which defines $\Lambda$ up to an arbitrary function of null Dalembertian.
In the five dimensional space, $B^{'A}$ has three transverse degrees of freedom. 
Indeed, the longitudinal
degree of freedom can be eliminated by a further suitable gauge transformation \re{gaugetf}
of null Dalembertian.
 
In the main part of the text, we will use,
for the usual case $\lambda=0$, 
the equations \re{square5} and \re{gauge5} for all $M^2$.

\vskip 0.3 cm
\subsubsection{Equations of motion. Generalized case ($\lambda\neq 0$) \label{massiveeqgen}}
\vskip 0.3 cm

\noindent The equations of motion in the generalized case \re{lag2}-\re{lambdadef},
 corresponding to 
$\lambda\neq 0$, are
\be
(\square_5+M^2)B^A+(\lambda-1)\,\partial^A(\partial_B B^B)=0.
\label{square5b}
\ee

\vskip 0.3 cm
\noindent{\bf{Generalized Case ($\lambda\neq 0$) with $M^2\neq 0$}} 
\vskip 0.3 cm
\noindent For $\lambda\neq 0$ and $M^2\neq 0$, defining the new vector field $B^{'A}$ 
and the new scalar field $B'$ by 
\bea
(B')^A&=&B^A+\frac{\lambda}{M^2}\partial^A\left(\partial_LB^L \right) 
    \nonumber\\
B'&=&\partial_A B^A \ ,
\label{newfielddef}
\eea
one easily sees that they are the physical fields  
as they satisfy, from \re{square5b}, 
the usual physical equations 
\bea
\left(\square _{5}+M^2\right){B'}^{A}&=&0
    \nonumber\\
\partial _{A}{B'}^{A}&=&0
    \nonumber\\
\left(\square _{5}+\frac{M^2}{\lambda}\right)B'&=&0.
\label{caselambda}    
\eea
Let us stress that the bulk mass squared of the 
vector field is $M^2$ while the bulk mass squared of 
the scalar field is multiplied by $1/\lambda$.

Inversely, if $B^{'A}$ and $B'$ are given such that they satisfy \re{caselambda}, 
then $B^A$ defined by
\be
B^A=B^{'A}-\frac{\lambda}{M^2}\partial^A B'
\label{inverseB}
\ee
satisfies the initial equation \re{square5b}. This shows that the physical
fields $B^{'A},B'$ and the initial fields $B^{A}$ are equivalent
ways of describing the theory.

\vskip 0.3 cm
\noindent{\bf{Generalized Case ($\lambda\neq 0$) with $M^2= 0$}} 
\vskip 0.3 cm
\noindent 
For $M^2=0$, the equations of motion in the generalized case ($\lambda\neq 0$) reduce to
\be
\square_5 B^A+(\lambda-1)\,\partial^A(\partial_B B^B)=0.
\label{square5bm0}
\ee
Let us note that the full gauge invariance (Eq.\re{gaugenorm} with $\Lambda$ arbitrary) 
of the usual Lagrangian ($\lambda=0$) is lost, but
there is a remaining ``partial gauge'' invariance
provided that
\be
\square_5 \Lambda=0.
\label{remaintf}
\ee

It is convenient to define the new fields $B^{'A}$ and $B'$ by
\bea
(B')^A&=&B^A+\partial^A\Sigma
    \label{gaugetfm0a}\\
B'&=&\partial_A B^{A}
\label{gaugetfm0}
\eea
with $\Sigma$ chosen to satisfy
\be
\square_5 \Sigma=\left(\lambda-1\right)\partial_A B^A.
\label{lambdam0}
\ee
Remark that $\Sigma$ is determined up to a function of null Dalembertian
i.e. $(B')^A$ can still be changed by a partial gauge transformation.

The new fields then obey the equations
\bea
\square_{5}{B'}^{A}&=&0
    \label{caselambdam0a}\\
\square _{5}B'&=&0
    \label{caselambdam0b}\\
\partial_A B^{'A}&=&\lambda B'.
\label{caselambdam0c}    
\eea
The interpretation of $B'$ as a massless physical scalar field poses no problem.
The interpretation of $B^{'A}$ as a massless vector field is more delicate 
as the Lorenz gauge condition
is missing. In momentum space (with $p^2=0$ by \re{caselambdam0a}), the five degrees of freedom of $B^{'A}$
decompose as follows. The divergence part $\partial_A B^{'A}$ is  
equivalent to the scalar $B'$ by \re{caselambdam0c}.
Any longitudinal part (in the direction of $p^{A}$)
is physically irrelevant as it can be removed by a suitable 
partial gauge transformation \re{remaintf}. 
The truly transverse part 
with three degrees of freedom describes a 
physical massless vector field.
The transverse part is defined up to an arbitrary longitudinal part i.e. 
belongs to the equivalence 
class, in the mathematical sense, of fields differing by partial gauge transformations.

In the main part of the paper, a complete discussion of the usual case
will be given. The changes which follow 
from the generalized case ($\lambda\neq 0$)
are outlined. 


\subsection{Basic boundary relations \label{flatBC}}

\vskip 0.3 cm
\subsubsection{Basic boundary relations. Usual case ($\lambda= 0$) \label{flatBCgene0}}
\vskip 0.3 cm

\noindent 
In order to guarantee the reality of the bulk mass squared $M^2$ 
(see \re{square5} and  \re{square5m0}), 
following the approach
discussed and justified at great length in our preceding articles, 
\cite{GN1}, 
we impose  that 
the operator $\square_5$ is an observable and hence hermitian
(more precisely symmetric)  
\bea
(C,{\rm{O}} B)&=&({\rm{O}} C,B)
    \label{herm0}\\
{\rm{O}}=\square_5        
    \label{herm1}
\eea
for the invariant scalar product of real vector fields
\begin{equation}
(C,B)=\int_{-\infty}^{+\infty}\!\!\!\!\!d^4\bx\int_{0}^{2\pi R} \!\!\!\!\!ds\,\,
      C_A(\bx,s)B^A(\bx,s). 
\label{scalprod1}
\end{equation}
This leads to 
\be
\int \!\!d^5x \,\,\partial_A\!\biggl(C_B(\partial^A B^B)-(\partial^AC_B)B^B\biggr)=0
\label{BCrelf}
\ee
which, taking into account the decrease
of the fields at
infinity in the $\bx$ space
(essentially faster than $1/\sqrt{\mid\bx\mid}$), becomes
the basic boundary relation
\be
\int d^4\bx \ \biggl[(\partial_s C_A)\, B^A-C_A \,(\partial_s B^A)\biggr]^{s=2\pi R}_{s=0}=0.
\label{BCrel0}
\ee
In the next section, upon the Kaluza-Klein reduction, 
the boundary conditions for the reduced fields 
will be obtained from reduced boundary relations compatible with \re{BCrel0}. 

\vskip 0.3 cm
\subsubsection{Basic boundary relations. Generalized case ($\lambda\neq 0$) \label{flatBCgene}}
\vskip 0.3 cm

\noindent In the generalized case, the equations of motion \re{square5b} are written
\be
\Omega^A_{\phantom{A}B} B^B=-M^2 B^A
\label{square5bb}
\ee
with the operator 
\be
\Omega^A_{\phantom{A}B}=\square_5\delta^A_B+\left(\lambda-1\right)\partial^A\partial_B.
\label{Omega}
\ee
In order for the eigenvalue $M^2$ to be real, the hermiticity of this operator is imposed
\be
     (\Omega C,B)=(C,\Omega B)
\label{Omegaherm}
\ee
for the scalar product \re{scalprod1}.
This leads to the boundary relation
\bea
&&{{\int}} d^4 \bx 
\ \biggl[\biggl(C_A\left(\partial_s B^A\right)-\left(\partial_s C_A\right)B^A\biggr)\biggr.
      \nonumber\\
&&\quad\quad\quad \biggl.+\left(\lambda-1\right)\biggl(C_S\left(\partial_A B^A\right)-\left(\partial_A C^A \right)B_S\biggr)
\biggr]^{2\pi R}_0
=0
\label{BCgene}
\eea
which generalizes \re{BCrel0}, since for $\lambda=0$,
$\partial_A B^A=0$ by \re{gauge5}).


\subsection{Kaluza-Klein reduction. Mass towers. Boundary conditions \label{flatKKreduc}}

\vskip 0.3 cm
\subsubsection{Kaluza-Klein reduction. Usual case ($\lambda= 0$) \label{KKreducgen0}}
\vskip 0.3 cm

\noindent We perform the following Kaluza-Klein reduction
\bea
B^{\mu}( \bx,s)&=&\sum _{r}B_r^{[V]}(s)\,\lambda_r^{\mu}(\bx)
   \label{a1}\\
B^{5}( \bx,s)&=&\sum _{p} B_p^{[S]}(s)\, \phi_p(\bx)
\label{a2}
\eea
which breaks the SO(4,1) covariance 
and respects the SO(3,1) covariance in any brane (at fixed $s$).

The functions $B_r^{[V]}(s)$ and $B_r^{[S]}(s)$ \re{a1},\re{a2} are supposed to satisfy
\bea
\partial_s^2\, B_r^{[V]}(s)&=&-\left(v^{[V]}_r\right)^2\, B_r^{[V]}(s)
  \label{eqbx}\\
\partial_s^2\, B_p^{[S]}(s)&=&-\left(v^{[S]}_p\right)^2\, B_p^{[S]}(s)
  \label{eqbs}
\eea
in such a way that the basic equations \re{square5}, applied for
each $r$ (or $p$), reduce to
\bea
\left(\square_4 +\left(m^{[V]}_r\right)^2\right)\lambda_r^{\mu}(\bx)&=&0
   \label{eqBmu}\\
\left(\square_4 +\left(m^{[S]}_p\right)^2\right)\phi_p(\bx)&=&0.
   \label{eqBs}
\eea    
The 4-dimensional masses for 
the 4-vector $\lambda_r^{\mu}(\bx)$ and for the 4-scalar 
$\phi_p(\bx)$  are then given by
\bea
\left(m^{[V]}_r\right)^2&=&\left(v^{[V]}_r\right)^2+M^2
     \nonumber\\
\left(m^{[S]}_p\right)^2&=&\left(v^{[S]}_p\right)^2+M^2.
\label{mass}
\eea

The indices $r$ and $p$ in \re{a1} and \re{a2} refer 
respectively to the 
4-dimensional masses squared $\left(m^{[V]}_r\right)^2$ and $\left(m^{[S]}_p\right)^2$ with by convention 
$\left(m^{[V]}_r\right)^2$ increasing with $r$ and $\left(m^{[S]}_p\right)^2$ increasing with $p$.
Together they build up the $B^A$ tower. 

Imposing the hermiticity \re{herm0} of the operator $\partial^2_s$ appearing in \re{eqbx}, \re{eqbs}
for the scalar product between functions of $s$
\be
(G,F)=\int_{s=0}^{2\pi R} G(s)F(s)\, ds 
\label{scalprod2}
\ee
ensures the reality of its eigenvalues and so of the corresponding 4-dimensional masses squared
(which could be negative).
This leads to boundary 
relations for the reduced functions $B_r^{[V]}$, $B_p^{[S]}$ (defined from $B^{A}$ in \re{a1},\re{a2}),
$C^{[V]}_m$ and $C^{[S]}_q$ (defined similarly from $C^A$)
\bea
  \biggl[(\partial_s C_m^{[V]})\, B_r^{[V]}-C_m^{[V]} \,(\partial_s B_r^{[V]})\biggr]^{2\pi R}_0&=&0
 \label{BCrel1}
\\
  \biggl[(\partial_s C_q^{[S]})\,B_p^{[S]}-C_q^{[S]}\,(\partial_s B_p^{[S]})\biggr]^{2\pi R}_0&=&0.
\label{BCrel2}
\eea
These boundary relations are compatible with the basic boundary relations \re{BCrel0}.

Boundary conditions result from these boundary relations
under the hypothesis that, on one side,
all the reduced fields related to $B^{\mu}$ and $C^{\mu}$ 
live in some common domain in the Hilbert space and that,
on the other side, all the reduced fields in $B^{5}$ and $C^{5}$ live in some,
eventually different from the first one,
common domain. 
Hence, the reduced fields
are submitted to boundary conditions
which must be independent of the index $r$ in \re{a1}, independent of the index $p$ 
in \re{a2}, but may depend on 
the $[V]$ or $[S]$ index. 

Thus, for any field $B$ the boundary conditions 
have to be expressed by two linear homogeneous relations, 
independent of $r$, among
the following quantities
\be
B_r^{[V]}(0)\ ,\ B_r^{[V]}(2\pi R)\ ,\ \partial_s B_r^{[V]}(0)\ ,\ \partial_s B_r^{[V]}(2\pi R)
\label{bxcond1}
\ee
and similarly by two such relations, independent of $p$, among
\be
B_p^{[S]}(0)\ ,\ B_p^{[S]}(2\pi R)\ ,\ \partial_s B_p^{[S]}(0)\ ,\ \partial_s B_p^{[S]}(2\pi R).
\label{bscond1}
\ee
For $B_n^{[S]}$, all the sets of allowed boundary conditions 
compatible with the 
boundary relations \re{BCrel2} are listed 
in Table \re{tablebounds}. 
The boundary conditions for $B_n^{[V]}$
are identical to those for $B_n^{[S]}$ with the parameters $\{\alpha_1^{[S]},\dots\}$ 
replaced by
$\{\alpha_1^{[V]},\dots\}$. Remark that all these boundary conditions are essentially the same as those 
allowed for a free real scalar field 
(see the first reference in \cite{GN1}).

The 
eigenvalues $(v_r^{[V]})^2$ of Eq.\re{eqbx} corresponding to a chosen set of boundary conditions 
for $B_r^{[V]}$
define the corresponding Kaluza-Klein mass tower
for $B^{\mu}$, and analogously  the $(v_p^{[S]})^2$ define the mass tower for $B^{S}$.
The full $B^A$ tower is composed 
of all the 
vector $(m_r^{[V]})^2$ and scalar $(m_p^{[S]})^2$ squared masses \re{mass} organized in such a way that
$m_{\{n\}}^2$ increases with some index which we call $\{n\}$.

There are two possibilities. Either the same set of boundary conditions
is imposed to $B_r^{[V]}$ and to $B_p^{[S]}$, in which case the corresponding spectra 
\re{eqbx},\re{eqbs} of eigenvalues 
coincide. Or different sets of boundary conditions are imposed to $B_r^{[V]}$ and to $B_p^{[S]}$,
in which case the Kaluza-Klein tower is composed of the two corresponding spectra. This conforms to the idea
that SO(4,1) is broken to SO(3,1) both by the compactification of the $s$ coordinate and by the
Kaluza-Klein reduction \re{a1}-\re{a2}.
In general, for a given $m_{\{n\}}^2$, there are three possible situations 
\begin{description}
\item{(1)} a vector ($B^{[V]}_{\{n\}}\lambda^{\mu}_{\{n\}}\neq 0$) and no scalar particle
( $B^{[S]}_{\{n\}}\phi_{\{n\}}=0$)\label{S1}

\item{(2)} no vector ($B^{[V]}_{\{n\}}\lambda^{\mu}_{\{n\}}= 0$) and a scalar particle
($B^{[S]}_{\{n\}}\phi_{\{n\}}\neq 0$)\label{S2}

\item{(3)} a vector ($B^{[V]}_{\{n\}}\lambda^{\mu}_{\{n\}}\neq 0$) and a scalar particle
($ B^{[S]}_{\{n\}}\phi_{\{n\}}\neq 0$)\label{S3}.

\end{description}
 Depending on ${\{n\}}$, any of these three situations may occur.

\vskip 0.3 cm
\subsubsection{Kaluza-Klein reduction. Generalized case ($\lambda\neq 0$)\label{KKreducgen}}
\vskip 0.3 cm

\noindent In the generalized case $\lambda\neq 0$, 
starting from the equations of motion 
\re{caselambda} or \re{caselambdam0a}-\re{caselambdam0c} 
for the
physical fields $B^{'A}$ and $B'$, 
we perform the following Kaluza-Klein reduction
\bea
B^{'\mu}( \bx,s)&=&\sum _{r}B_r^{'[V]}(s)\,\lambda_r^{'\mu}(\bx)
   \label{gena1}\\
B^{'5}( \bx,s)&=&\sum _{p} B_p^{'[S]}(s)\, \phi^{'}_p(\bx)
\label{gena2}\\
B^{'}( \bx,s)&=&\sum _{v} B_v^{'}(s)\, \psi^{'}_v(\bx).
\label{gena3}
\eea
Imposing the hermiticity of $\partial^2_s$ and
following the usual procedure, one obviously finds that the reduced fields
$B_r^{'[V]}(s)$, $B_p^{'[S]}$ and $B_v^{'}(s)$ must satisfy boundary relations 
of the form \re{BCrel1}-\re{BCrel2}
\bea
  \biggl[(\partial_s C_m^{'[V]})\, B_r^{'[V]}-C_m^{'[V]} \,(\partial_s B_r^{'[V]})\biggr]^{2\pi R}_0&=&0
 \label{BCrelgen1}
\\
  \biggl[(\partial_s C_q^{'[S]})\,B_p^{'[S]}-C_q^{'[S]}\,(\partial_s B_p^{'[S]})\biggr]^{2\pi R}_0&=&0
\label{BCrelgen2}
\\
  \biggl[(\partial_s C_v^{'})\,B_w^{'}-C_v^{'}\,(\partial_s B_w^{'})\biggr]^{2\pi R}_0&=&0.
\label{BCrelgen3}
\eea
The derived boundary conditions must be the same for all 
the reduced components of each of the fields. In other words, the boundary conditions
in terms of the fields $B_r^{'[V]}(s)$, $B_p^{'[S]}$ and $B_v^{'}(s)$ evaluated at $s=0$ and $s=2\pi R$
(see the discussion of \re{bxcond1})
must be independent of the indices $r$, $p$ and $v$ respectively.

Refering to \re{inverseB}, it is natural, 
in order to obtain definite boundary conditions for $B^A$, 
to impose the same boundary conditions for $B^{'A}$ and 
$\partial^A B'$
\bea
B^{'[V]}_r(s)\quad &{\rm{same\ boundary\ conditions\ as}}&\quad \ \ B^{'}_v(s)
   \label{equivBC1}\\
B^{'[S]}_r(s)\quad &{\rm{same\ boundary\ conditions\ as}}&\quad \partial_s B^{'}_v(s)
\label{equivBC2}
\eea	
independently of $r$ and $v$.

\vskip 0.3 cm
\subsubsection{Physical implication ($\lambda\neq 0$)\label{Physimp}}
\vskip 0.3 cm

\noindent Since
$B^{'[V]}_r$ and $B^{'}_v$ are submitted to the same boundary conditions 
\re{equivBC1}, their eigenvalue spectrum $v_n^2$ are identical.
Hence, the scalar and vector mass towers of the $\lambda\neq 0$ case are 
the scalar and vector mass towers of the $\lambda= 0$ case accompanied 
by a new scalar mass tower (with fields $\psi^{'}_p(\bx)$) with all their masses, 
according to \re{caselambda}, shifted by
\be
{\rm{Square\ mass\ shift\ }}=(m_n^{[\psi]})^2-(m_n^{[V]})^2=\frac{1-\lambda}{\lambda}M^2.
\label{massshift}
\ee


\subsection{Gauge condition \label{flatKKeq}}

\vskip 0.3 cm
\subsubsection{Gauge condition. Usual case ($\lambda= 0$) \label{KKgen0}}
\vskip 0.3 cm

\noindent The Lorenz condition \re{gauge5}, restricted to one $\{n\}$, reduces to
\be
B_{\{n\}}^{[V]}\,\left(\partial_{\mu}\lambda_{\{n\}}^{\mu}\right)+\left(\partial_s B_{\{n\}}^{[S]}\right)\,\phi_{\{n\}}=0.
\label{eqgauge}
\ee
In order to have a non-zero vector and/or a non-zero scalar part $\{n\}$ in the tower
(not both zero), 
this equation implies in general (taking into account an arbitrary normalisation) 
one of the following four distinct physical cases
\bea
&&{\rm{Generic\ Case\ 1}}
       :\ B_{\{n\}}^{[V]}\neq 0 \ \,,
       \ \partial_s B_{\{n\}}^{[S]}=B_{\{n\}}^{[V]}\ ,
       \left\{
       \begin{array}{l}
       \partial_{\mu}\lambda_{\{n\}}^{\mu}+\phi_{\{n\}}=0\\
       \phi_{\{n\}}\neq 0
       \end{array}
       \right. 
   \label{connectgen1}\\
&&{\rm{Generic\ Case\ 2}}
       :\ B_{\{n\}}^{[V]}\neq 0 \ \,,
       \ {\rm{no\ scalar}} \  \quad \quad\ ,
       \quad\ \partial_{\mu}\lambda_{\{n\}}^{\mu} =0
   \label{connectgen2}\\
&&{\rm{Special\ Case\ 1}}
     :\ B_{\{n\}}^{[V]}\neq 0\ \,,\ B_{\{n\}}^{[S]}=1 
      \quad\quad\ \ \,,
      \left\{
       \begin{array}{l}
       \partial_{\mu}\lambda_{\{n\}}^{\mu}=0\\
       \phi_{\{n\}}\neq 0
       \end{array}
       \right. 
   \label{connect1}\\
&&{\rm{Special\ Case\ 2}}
    :\ {\rm{no\ vector}}\,,\ B_{\{n\}}^{[S]}=1 \quad\quad\ \ \,,\quad\ \phi_{\{n\}}\neq 0.                   
   \label{connect2}
\eea  

From these equations, the physical content of the tower at level $\{n\}$ can be easily deduced
for every case \re{connectgen1}-\re{connect2}. In general, the Generic Cases  
are related to tower states
while the Special Cases correspond 
(within a tower) to specific states characterized by $v_{\{n\}}^2=0$.

\begin{description} 

\item 

{\bf{$\bullet$ Generic Case 1 when}} $m_{\{n\}}^2=M^2+v_{\{n\}}^2\neq 0$ (see \re{connectgen1}). 
For any such $\{n\}$, there is a 4-dimensional vector field $A^{\mu}_{\{n\}}$ 
and a 4-dimensional scalar field $\phi_{\{n\}}$ given by 
\be
\hspace{-0.7cm}
{
{\rm{Generic\ Case\ 1}}:
\left\{
\begin{array}{c}
{\rm{vector\ field\ }} A^{\mu}_{\{n\}} \cr
{\rm{scalar\ field\ }} \phi_{\{n\}}\cr  
m_{\{n\}}^2 \neq 0 
\end{array}
\right.
\left\{
\begin{array}{l}
A^{\mu}_{\{n\}}=\lambda_{\{n\}}^{\mu} - \frac{\partial^{\mu}\phi_{\{n\}}}{m_{\{n\}}^2}
   \cr
(\square_4+m_{\{n\}}^2 )\phi_{\{n\}}=0  
  \cr
(\square_4+m_{\{n\}}^2 )A^{\mu}_{\{n\}}=0 
  \cr
\partial_{\mu}A^{\mu}_{\{n\}}=0
\end{array}
\right. .
}
\label{mass4a}  
\ee
Note that the vector $A^{\mu}_{\{n\}}$ 
is a true physical vector field, 
which satisfies as expected the 4-dimensional Lorenz condition.

\item

{\bf{$\bullet$ Generic Subcase 1
when}}, for a given $\{n\}$, $m_{\{n\}}^2=M^2{+}v_{\{n\}}^2= 0$. There is 
then a 4-dimensional massless scalar field $\phi_{\{n\}}$, 
and a 4-dimensional massless vector field $A^{\mu}_{\{n\}}$
which satisfy  
\be
\hspace{-0.8cm}
{
{\rm{Generic\ Subcase\ 1}}: 
\left\{
\begin{array}{c}
{\rm{\ vector\ field\ }} A_{\{n\}}^{\mu}\cr
{\rm{scalar\ field\ }}\phi_{\{n\}}\cr 
(m_{\{n\}}^A)^2=(m_{\{n\}}^{\phi})^2=0  
\end{array}
\right.
\left\{
\begin{array}{l}
A^{\mu}_{\{n\}}=\lambda_{\{n\}}^{\mu}  \cr   
\square_4 A^{\mu}_{\{n\}}=0 \cr
\square_4 \phi_{\{n\}}=0  \cr
\partial_{\mu}A^{\mu}_{\{n\}}=-\phi_{\{n\}}
\end{array}
\right.
}.
\label{mnzeroa}
\ee
The equations for $A^{\mu}_{\{n\}}$ and $\phi_{\{n\}}$ in four dimensions are of the same form as 
the equations \re{caselambdam0a}-\re{caselambdam0c} for respectively $B^{'A}$ and $B'$ 
in five dimensions for $\lambda=-1$. They belong to a generalized 4-dimensional case
with $\lambda\neq 0$.
Refering to the discussion following \re{caselambdam0c}, 
we conclude that $\phi_{\{n\}}$ describes a scalar massless particle in the brane and that
$A^{\mu}_{\{n\}}$ describes a particle 
whose 4-dimensional divergence is proportional to $\phi_{\{n\}}$, whose longitudinal part 
can be removed by a suitable partial gauge transformation and whose 
two truly transverse degrees of freedom describe a physical massless vector field in the brane.

\item 

{\bf{$\bullet$ Generic Case 2 with}} $m_{\{n\}}^2=M^2+v_{\{n\}}^2$ (see \re{connectgen2}). 
For any $\{n\}$
with $m_{\{n\}}^2=v_{\{n\}}^2+M^2$, there is a true 4-dimensional vector field $A_{\{n\}}^{\mu}$   
and no scalar field.
\be
\hspace{-0.5cm}
{
{\rm{Generic\ Case\ 2}}: 
\left\{
\begin{array}{c}
{\rm{vector\ field\ }}A_{\{n\}}^{\mu} \cr
{\rm{no\ scalar\ field}}\cr
m_{\{n\}}^2=v_{\{n\}}^2+M^2
\end{array}
\right.
\left\{
\begin{array}{l}
A_{\{n\}}^{\mu}=\lambda_{\{n\}}^{\mu}\cr
(\square_4 +m_{\{n\}}^2)A_{\{n\}}^{\mu}=0\cr
\partial_{\mu}A_{\{n\}}^{\mu}=0\cr
{\rm{no\ }}\phi_{\{n\}}
\end{array}
\right.
.   
}
\label{mass1}
\ee

\item

{\bf{$\bullet$ Special Case 1}} \re{connect1}. Both a 4-dimensional vector $A_{\{n\}}^{\mu}$ 
and a 4-dimensional scalar 
$\phi_{\{n\}}$ field are present in the brane and, from \re{eqbs},
$v_{\{n\}}^2=0$. Their mass squared is equal to the bulk mass squared.
\be
\hspace{-0.5cm}
{
{\rm{Special\ Case\ 1}}: 
\left\{
\begin{array}{c}
{\rm{vector\ field\ }} A_{\{n\}}^{\mu}\cr
{\rm{scalar\ field\ }} \phi_{\{n\}} \cr
m_{\{n\}}^2=M^2    
\end{array}
\right.
\left\{
\begin{array}{l}
A_{\{n\}}^{\mu}=\lambda_{\{n\}}^{\mu}
  \cr
(\square_4+M^2 )\phi_{\{n\}}=0  
  \cr
(\square_4+M^2 )A_{\{n\}}^{\mu}=0 
  \cr
\partial_{\mu}A_{\{n\}}^{\mu}=0
\end{array}
\right. .
}
\label{mass2}
\ee

\item 

{\bf{$\bullet$ Special Case 2}} \re{connect2}. There is a no 
4-dimensional vector state but a lonely 4-dimensional scalar state $\phi_{\{n\}}$
is present and, from \re{eqbs},
$v_{\{n\}}^2=0$. This means that the scalar has a brane mass squared equal to the bulk mass
squared.
\be
{\rm{Special\ Case\ 2
}}: 
\left\{
\begin{array}{c}
{\rm{no\ vector\ field}}\cr
{\rm{scalar\ field\ }} \phi_{\{n\}}\cr
m_{\{n\}}^2=M^2   
\end{array}
\right. 
\left\{
\begin{array}{l}
(\square_4+M^2 )\phi_{\{n\}}=0\cr
{\rm{no\ }} \lambda_{\{n\}}^{\mu}  
\end{array}
\right. .
\label{mass3}
\ee

\end{description}

\vskip 0.3 cm
\subsubsection{Gauge condition. Generalized case ($\lambda\neq 0$) \label{KKgen}}
\vskip 0.3 cm

\noindent Remembering the discussion concerning 
Generic Case 1 \re{connectgen1}
and Generic Case 2 \re{connectgen2} and taking into account Eq.\re{inverseB}, \re{equivBC1},
\re{equivBC2}, we have the 
general possibilities
\bea
\hspace{-1.5cm}&&{\rm{Generalized\ Generic\ Case\  1:\ }}
    {\rm{same\ BC\ for\ }} B^{'[V]}_r,\ \partial_s B^{'[S]}_r {\rm{\ and\ }} B^{'}_v     
   \label{connectext1}\\ 
\hspace{-1.5cm}&&{\rm{Generalized\ Generic\ Case\ 2:\ }}
   B^{'[S]}_r=0,\     {\rm{same\ BC\ for\ }} B^{'[V]}_r{\rm{\ and\ }} B^{'}_v.     
   \label{connectext2}
\eea  
The discussion of the Lorenz condition \re{eqgauge} (for $B^{'[V]}$ and 
$B^{'[S]}$) and of the
the Extended Special Cases 1 and 2 follow the same line as in the $\lambda= 0$ case 
with obvious changes.
From here on, we concentrate on the usual case ($\lambda=0$).


\subsection{Final sets of allowed boundary conditions \label{flatBCfinal}}

The sets of boundary conditions (abbreviated BC in this subsection)
for $B_{\{n\}}^{[S]}$ and $B_{\{n\}}^{[V]}$ (see Table \re{tablebounds})
have to be made compatible with each other, taking into account
the eigenvalue equations \re{eqbx},\re{eqbs}
and in particular the Lorenz condition \re{eqgauge}.
The results, as obtained in App.\re{joinbound}, are summarized hereafter.
The generic cases are related to 
BC which lead to towers composed of an infinite number of states.
The special cases are related to BC which lead to the presence of specific states
in the towers.

\vskip 0.3 cm
\noindent{\bf{$\bullet$ Generic Case 1 \label{genericas1}}}
\vskip 0.3 cm

\noindent In the Generic Case 1 \re{connectgen1}, the 
allowed sets of BC
for  $B_n^{[V]}$ and $B_n^{[S]}$
are given in 
the following Table

\be
\hspace{-0.2 cm}
{\rm{Sets\ G1\ }}\left\{ \ \  
{\tiny{
\begin{tabular}{|c|c|c|c|c|}
\hline
$\phantom{\Biggl[\Biggr]}$
Set
$\phantom{\Biggl[\Biggr]}$
     &               &
     &               &
     \\
\hline \hline
$\phantom{\Biggl[\Biggr]}$
G1a
$\phantom{\Biggl[\Biggr]}$
     &  $ B_n^{[V]}(2\pi R)$  
     &  $\partial_s B_n^{[V]}(2\pi R)$ 
     &  $ B_n^{[S]}(2\pi R)$ 
     &  $\partial_s B_n^{[S]}(2\pi R)$
     \\ 
     &  = &= &= & = 
     \\    		
$\phantom{\Biggl[\Biggr]}$
     & $ \alpha B_n^{[V]}(0)$  
     & $\frac{1}{\alpha}\partial_s B_n^{[V]}(0)$ 
     & $\frac{1}{\alpha} B_n^{[S]}(0)$ 
     & $\alpha\partial_s B_n^{[S]}(0)$ 	     
    \\ \hline
$\phantom{\Biggl[\Biggr]}$
G1b  
$\phantom{\Biggl[\Biggr]}$
&  $ B_n^{[V]}(0)=0$  
     &  $B_n^{[V]}(2\pi R)=0$ 
     &  $\partial_s B_n^{[S]}(0)=0$ 
     &  $\partial_s B_n^{[S]}(2\pi R)=0$
    \\ \hline  		
$\phantom{\Biggl[\Biggr]}$
G1c  
$\phantom{\Biggl[\Biggr]}$
     &  $ \partial_s B_n^{[V]}(0)=0$  
     &  $\partial_sB_n^{[V]}(2\pi R)=0$ 
     &  $B_n^{[S]}(0)=0$ 
     &  $B_n^{[S]}(2\pi R)=0$
    \\ \hline  		
$\phantom{\Biggl[\Biggr]}$
G1d  
$\phantom{\Biggl[\Biggr]}$
     &  $ B_n^{[V]}(0)=0$  
     &  $\partial_s B_n^{[V]}(2\pi R)=0$ 
     &  $\partial_s B_n^{[S]}(0)=0$ 
     &  $B_n^{[S]}(2\pi R)=0$
    \\ \hline  		
$\phantom{\Biggl[\Biggr]}$
G1e  
$\phantom{\Biggl[\Biggr]}$
     &  $ \partial_s B_n^{[V]}(0)=0$  
     &  $ B_n^{[V]}(2\pi R)=0$ 
     &  $ B_n^{[S]}(0)=0$ 
     &  $ \partial_s B_n^{[S]}(2\pi R)=0$
    \\ \hline 
\end{tabular} 
\ \ 
}} 
\right. . 
\label{caseG1}  
\ee 

\vskip 0.0 cm

\vskip 0.3 cm
\noindent{\bf{$\bullet$ Generic Case 2 \label{genericas2}}}
\vskip 0.3 cm

\noindent In the Generic Case 2 \re{connectgen2}, the allowed sets of BC are
\be
{\rm{Sets\  G2}} \left\{
               \begin{array}{ll}
           B_{\{n\}}^{[S]}       =0&{\rm{special\ solution\ compatible\ with}}       \cr
                                         &{\rm{any\ set\ of \ BC\ of\ Table\ \re{tablebounds}}}\cr           
           {\rm{for\ }}B_{\{n\}}^{[V]}\neq 0 & {\rm{any\ set\ of\ BC
                         \ of\ Table\ \re{tablebounds}}}\cr   
               \end{array}
               \right.    
.
        \label{caseG2}         
\ee

\vskip 0 cm
\vskip 0.3 cm
\noindent{\bf{$\bullet$ Special Case 1 \label{specialcas1}}}
\vskip 0.3 cm

\noindent In the Special Case 1 \re{connect1}, the allowed sets of BC are
\be
\hspace{-0.1cm}
{
{\rm{Set\  S1a}} \left\{
               \begin{array}{ll}
           B_{\{n\}}^{[S]}                =1&{\rm{special\  solution\ of \ a\ subset}}       \cr
                                         &{\rm{of\ the\ set            
                                         \ A1\ of\ Table\ \re{tablebounds}}}\cr
            & \quad
           \left\{
           \begin{array}{l}
            B_{\{n\}}^{[S]}(2\pi R)=B_{\{n\}}^{[S]}(0)+\alpha_2^{[S]} \partial_s B_{\{n\}}^{[S]}(0)        \cr
           \partial_s B_{\{n\}}^{[S]}(2\pi R)= \partial_s B_{\{n\}}^{[S]}(0)        
           \end{array}\right.\cr                                        
          {\rm{for\ }}B_{\{n\}}^{[V]}\neq 0 & {\rm{any\ set\ of\ BC
                                \ of\ Table\ \re{tablebounds}}}\cr 
               \end{array}
               \right.    
   }
        \label{caseS1a}         
\ee
\be
\hspace{-4.1 cm}{
{\rm{Set\  S1b}} \left\{
               \begin{array}{ll}
           B_{\{n\}}^{[S]}                =1&{\rm{special\  solution\ of\ a\ subset}}       \cr          
                         &{\rm{\ of\ the\ set\ A2\ of\ Table\ \re{tablebounds}}}\cr 
             &\quad\left\{
             \begin{array}{l}            
            \partial_s B_{\{n\}}^{[S]}(0)=0        \cr
           \partial_s B_{\{n\}}^{[S]}(2\pi R)= 0        
           \end{array}\right.\cr
           {\rm{for\ }}B_{\{n\}}^{[V]}\neq 0 & {\rm{any\ set\ of\ BC\ of\ Table\ \re{tablebounds}}}\cr  
               \end{array}
               \right.    
   .}
        \label{caseS1b}         
\ee

\vskip 0.3 cm
\noindent{\bf{$\bullet$ Special Case 2 \label{specialcas2}}}
\vskip 0.3 cm

\noindent In the Special Case 2 \re{connect2}, the allowed sets of BC are
\be
\hspace{0.37 true cm}
{\rm{Set\  S2a}} \left\{
               \begin{array}{ll}
           B_{\{n\}}^{[S]}                =1&{\rm{special\  solution\ of\ the\ subset}} \cr
                                         &{\rm{\ of\ the\ set          
                         \ A1\ of\ Table\ \re{tablebounds}}}\cr 
            &\quad
            \left\{
            \begin{array}{l} 
            B_{\{n\}}^{[S]}(2\pi R)=B_{\{n\}}^{[S]}(0)+\alpha_2^{[S]} \partial_s B_{\{n\}}^{[S]}(0)        \cr
           \partial_s B_{\{n\}}^{[S]}(2\pi R)= \partial_s B_{\{n\}}^{[S]}(0)        
           \end{array}
           \right.
           \cr
           B_{\{n\}}^{[V]}= 0 & {\rm{special\ solution\ of\ any}}\cr
                         &{\rm{set\ of\ BC 
                         \ of\ Table\ \re{tablebounds}}}\cr  
               \end{array}
               \right.    
        \label{caseS2a}         
\ee

\be
\hspace{-3.3 cm}
{\rm{Set\  S2b}} \left\{
               \begin{array}{ll}
           B_{\{n\}}^{[S]}                =1&{\rm{special\  solution\ of \ the\ subset}}       \cr
                                         &{\rm{of\ the\ set            
                               \ A2\ of\ Table\ \re{tablebounds}}}\cr
           & \quad\left\{
            \begin{array}{l}                    
            \partial_s B_{\{n\}}^{[S]}(0)=0        \cr
           \partial_s B_{\{n\}}^{[S]}(2\pi R)= 0       
           \end{array}
           \right. \cr
           B_{\{n\}}^{[V]}= 0 & {\rm{special\ solution\ of\ any}}\cr
                         &{\rm{set\ of\ BC  
                         \ of\ Table\ \re{tablebounds}}}\cr  
               \end{array}
               \right.    
.
        \label{caseS2b}         
\ee


\section{Physical considerations \label{phys}}

\subsection{Physical discussion of the
boundary conditions. Closure to a circle \label{locality}}

If one looks at the boundary conditions of Table \re{tablebounds}, one sees that there are two
very different situations. For the Sets A2, A3, A4, A5, the boundary conditions are local. 
The values of the fields are related at the same point, either at 
$s=0$ or at $s=2\pi R$. 
The boundary conditions A1 are of a rather different 
kind as they connect values of the fields evaluated at
these two different points of the $s$ domain.
In this latter case, the field
explores in fact its full domain at once. This is tantamount to action at a distance
or to non locality. In the third article of \cite{GN1}, 
we noted that this was not in contradiction with quantum mechanics.

Under the A1 boundary conditions with $\alpha_2=\alpha_3=0$,
the strip can be closed 
into a circle by identifying the points $s=0$ and $s=2\pi R$.   
With $\alpha_1=\alpha_4=1$ 
this leads to the periodic 
boundary conditions, and with $\alpha_1=\alpha_4=-1$ 
to the antiperiodic boundary conditions. They are both of local type.

\subsection{Mass considerations \label{physmass}}

For completeness, we will give, in the next subsection, a description of
all the allowed mass equations from which the mass 
towers are derived. We remind that we are here restricted to 
a five dimensional flat space with a compactified fifth coordinate. 
In general, an evaluation of the low lying mass eigenvalues
give values of the form $1/R$ multiplied by a factor of order 1.
For mass eigenvalues of the order of 1 TeV, this requires
$R$ to be of the order $10^{-16}$cm.

This value
is far away from the $R\approx 10^{10}$cm originating 
from the relation 
\be
M_{Pl}^2\approx V_{\{n\}} M_{*}^{n+2}
\label{Plmstar}
\ee
between the Planck mass $M_{Pl}$, 
the fundamental scale $M_{*}$ and 
the volume $V_5=(2\pi R)^5$
under the hypothesis that Gauss
law is valid in the five dimensional space.              
This idea has been discussed at length in the litterature \cite{ADD},
but the results are still under investigation. 
As is well-known the problem 
has been cured elegantly by Randall and Sundrum 
\cite{RS1} by the use of 
warped spaces (see also \cite{GP}) as will be discussed in a subsequent publication.

\subsection{Mass considerations. Experimental aspects \label{physmassexp}}

The presence or absence of scalars states in the Kalauza-Klein towers
can roughly be summarized as follows

\begin{description}

\item{(1) $\lambda\neq 0$, Generic Case 1 \re{mass4a}} 

Accompanying the 4-dimensional vector tower there are two 4-dimensional scalar towers. 
The masses in one 
of the
scalar tower are identical to the masses in the vector tower while the masses in the second scalar 
tower are shifted from the masses in the vector tower by the fixed amount \re{massshift}.

\item{(2) $\lambda\neq 0$, Generic Case 2 \re{mass1}}

Accompanying the 4-dimensional vector tower there is one 4-dimensional scalar tower.
The masses in the scalar 
tower are shifted from the masses in the vector tower by the fixed amount \re{massshift}.

\item{(3) $\lambda= 0$, Generic Case 1 \re{mass4a}}

Accompanying the 4-dimensional vector tower there is one 4-dimensional scalar tower. 
The masses in the scalar 
tower are identical to the masses in the vector tower.

\item{(4) $\lambda= 0$, Generic Case 2 \re{mass1}}

There is one 4-dimensional vector tower and no 4-dimensional scalar tower.

\end{description}

An examination of the 
masses of the known 
bosons as given by the Particle Data Group \cite{PDG},
reveals at first sight that, to any set of observed vector (or pseudo-vector) particles, 
there is no sign of a related set of scalar
(or pseudo-scalar) particles either with identical masses or 
with masses squared shifted by a fixed amount.

Hence, if a Kaluza-Klein tower of vector mesons exists
and some of its low lying states already appear in the Particle Data Group tables, the theory
must belong to the case (4) above i.e. must correspond to a Lagrangian with $\lambda= 0$
while the boundary conditions must belong to the Generic Case 2.


\section{Towers \label{towers}}

In this section, we give the main properties of the vector and related scalar towers
which follow from specific choices of allowed boundary conditions (\re{caseG1} and \re{caseG2}).
The detailed towers are obtained
by following the procedure used in the first article of \cite{GN1}

\subsection{Towers in the Generic Case G1a \label{G1a}}

The general solution of $B_{\{n\}}^{[S]}(s)$ for $v_n^2 > 0$ is a superposition of sine and cosine functions 
(see \re{solsincos}). For $v_n^2< 0$ it is a superposition of hyperbolic sine and cosine functions 
while for $v_n= 0$ it is a linear function in $s$. It is easy to see that 
the hyperbolic solution is incompatible
with the G1a boundary conditions in Table \re{caseG1} for any $\alpha$, 
while the linear solution can only be a constant and is compatible with $\alpha=1$ only. We thus focuss of the sine and cosine solution. 

Writing the general solution $B_{\{n\}}^{[S]}(s)$ of \re{eqbs} (taking $v_{\{n\}}>0$) as
\be
B_{\{n\}}^{[S]}(s)=\sigma_{\{n\}}^{[S]} \sin(v_{\{n\}} s)+\tau_{\{n\}}^{[S]}\cos(v_{\{n\}} s),
\label{solsincos}
\ee
with $B_{\{n\}}^{[V]}(s)=\partial_s B_{\{n\}}^{[S]}(s)$, 
the set of boundary conditions G1a \re{caseG1} leads to 
the eigenvalue equation for $v_{\{n\}}$ 
\be
\cos(2\pi v_{\{n\}} R)=\frac{2}{\alpha+\frac{1}{\alpha}}
\label{vnga}
\ee
with
\be
\sin(2\pi v_{\{n\}} R)=\epsilon\frac{\alpha^2-1}{\alpha^2+1} \quad,\quad \epsilon^2=1.
\label{vngb}
\ee
If $\alpha^2\neq 1$, the field parameters $\sigma_{\{n\}}^{[S]}$ and $\tau_{\{n\}}^{[S]}$ are related by 
\be
\tau_{\{n\}}^{[S]}=-\epsilon \alpha\sigma_{\{n\}}^{[S]}
\label{tausig}
\ee
while if $\alpha^2= 1$ they are arbitrary.

Calling 
\be
w=\frac{1}{\pi }\arccos\left(\frac{2}{\alpha+\frac{1}{\alpha}}\right) \quad,\quad 0\leq w\leq 1
\label{gav}
\ee
the mass tower $m_{\{n\}}^2$ is composed of the $\epsilon=+1$ and $\epsilon=-1$ intertwined
branches 
\bea 
n=2p+1&,&\quad m_{\{2p+1\}}^2=M^2+\frac{\left(2p+w\right)^2}{4R^2}\quad,\quad p\geq 0
    \nonumber\\
n=2p\quad&,&\quad m_{\{2p\}}^2=M^2+\frac{\left(2p-w\right)^2}{4R^2} \quad\quad,\quad p\geq 1.
\label{gbmn1a}
\eea
Note that the index $\{n\}$ is defined in such a way that the 
first state in the tower corresponds to $n=1$ and that the masses are ordered in such a way
that they increase with $n$.
Remark also that for $\alpha=\pm 1$, every state in the tower is twice degenerate
except for $\alpha=1$ where the first state with $m_{\{1\}}^2$
is single.
Indeed, for $\alpha=1$ ($w=0$) one has $m_{\{2q\}}^2=m_{\{2q+1\}}^2, q\geq 0$ and 
for $\alpha=-1$ ($w=1$) one has $m_{\{2q+1\}}^2=m_{\{2(q+1)\}}^2,\  q>0$.

The only possibility for having $v_{\{1\}}=0$  
is a constant
$B_{\{n\}}^{[S]}(s)$. It implies $\alpha=1$
and
corresponds to the Special Case 1 \re{connect1} or Special Case 2 \re{connect2}
(see \re{mass2} or \re{mass3}).

The lowest mass squared in the tower \re{mass} is always larger than the bulk mass squared  $M^2$ 
if $\alpha\neq 1$
and equal to it if $\alpha=1$. Note that the bulk mass squared can a priori be negative
which means that the original five-dimensional field is tachyonic. 
This could lead, at the bottom of the tower, to the appearance of one or more four-dimensional tachyons. 
A zero mass state (eventually twice degenerate for $\alpha^2=1$) can occur provided the boundary parameter $\alpha$ 
is such that $M^2$ equals $-v_{\{n\}}^2$ i.e. the negative   
of one of the eigenvalues.

\subsection{Towers in the Generic Case G1b \label{G1b}}

The boundary conditions in the Case G1b of \re{caseG1},
when imposed
to the sine and cosine solution \re{solsincos} for
$B_{\{n\}}^{[S]}(s)$ ($v_{\{n\}}>0$), lead
to the well-known
regular tower 
\be
v_{\{n\}}=\frac{n}{2R}\quad {\rm{for}}\ \  n=1,2,\dots
\label{gbmn1b}
\ee
as the solutions of the equation
\be
\sin(2\pi R v_{\{n\}})=0.
\label{gbmn1beq}
\ee
There is no hyperbolic sine or cosine solution ($v_{\{n\}}^2<0$).
The case $v_{\{n\}}=0$, corresponding to a linear $B_{\{n\}}^{[S]}(s)$, leads to a constant field,
say $B_{\{n\}}^{[S]}(s)=1$ by normalisation.
This corresponds to the lowest state ($n=0$) in the tower
which is of the \re{connect1} form. The resulting values of the masses squared of all the
states in the tower 
are finally
\be
m_{\{n\}}^2=M^2+\left(\frac{n}{2R}\right)^2 \quad {\rm{for}}\ n=0,1,\dots\ .
\label{gbmn1bmass}
\ee

\subsection{Towers in the Generic Case G1c \label{G1c}}

Analogously to the preceding G1b case,
the boundary conditions in the Case G1c of \re{caseG1} lead 
for $B_{\{n\}}^{[S]}(s)$
to the sine and cosine solution when $v_{\{n\}}^2>0$ 
but neither to a constant solution for $v_{\{n\}}=0$ nor to a hyperbolic solution 
for $v_{\{n\}}^2<0$.
The same
regular tower \re{gbmn1bmass} results
except that the $n=0$ state is absent.

\subsection{Towers in the Generic Cases G1d, G1e \label{G1d}}
Analogously to the preceeding G1b, G1c cases,
the boundary conditions in the Case G1d and G1e of \re{caseG1} lead to the sine and cosine solution 
for $v_{\{n\}}^2>0$ 
but neither to a constant solution for $v_{\{n\}}^2=0$ nor to a hyperbolic solution for $v_{\{n\}}^2<0$ .
The $v_{\{n\}}>0$ must satisfy the equation
\be
\cos(2\pi R v_{\{n\}})=0
\label{gbmn1deq}
\ee
and the masses squared of the states in the tower are
\be
m_{\{n\}}^2=M^2+\frac{1}{4R^2}\left(n+\frac{1}{2}\right)^2 \quad {\rm{for}}\ n=0,1,\dots\ .
\label{gbmn1dmass}
\ee

\subsection{Towers in the Generic Case G2 \label{casegen2}}

Since the $B_{\{n\}}^{[V]}$ are allowed to satisfy 
any set of boundary conditions \re{caseG2} as
summarized in Table \re{tablebounds} 
(with $[P]$ replaced by $[V]$) 
and since $B_{\{n\}}^{[S]}=0$, there are only vector towers and
no corresponding scalar towers.  
The allowed towers are summarized in Table \re{tablescaltw}. They
are in complete correspondance 
with the towers obtained for scalar states in a flat space 
(see the first article in \cite{GN1}).
In general, one finds that the masses in the towers
are derived from equations involving trigonometric functions and 
polynomials. These equations depend in a specific way on the parameters 
defining the boundary conditions and lead to less regular towers than the usual 
linear forms \re{gbmn1bmass}, \re{gbmn1dmass} which were originally associated to 
Kaluza-Klein towers.


\section{Conclusions \label{conclusion}}

The occurence of Kaluza-Klein towers for vector fields 
propagating in a five dimensional compactified flat space
might be interesting to look at, at high energies.
In this paper, we have analyzed and presented the results of a careful study 
of all the sets of allowed boundary conditions which follow from the 
physical requirement of hermiticity (symmetry or self-adjointness) of the operators
whose eigenvalues are related to the Kaluza-Klein mass tower states, as seen in a 4-brane.  

Our approach starts from the most general free Lagrangian quadratic
in the bulk vector field and its derivatives. This Lagrangian depends on two free parameters, 
the bulk mass $M$ 
and one extra parameter $\lambda$ (see Eq.\re{lag2} and Eq.\re{lambdadef}).  
The sets of allowed 
boundary conditions are restricted by the all important Lorenz 5-dimensional gauge condition.

In the case of the generalized Lagrangian ($\lambda\neq 0$), 
there appears,
for any set of boundary conditions and apart from the main tower 
constituted by vector states (with possibly 
an associated tower of
scalar states with the same mass spectrum), 
an extra tower of scalar particles with masses squared which differ 
from the main tower masses squared
by a universal additive term $(1-\lambda)M^2/\lambda$ which does 
not depend on the parameters chosen for the boundary conditions. 

For the main tower,
there are essentially two distinct situations.
In the first one, the four-dimensional vector tower is associated to a 
four-dimensional scalar tower with coinciding masses.  Their spectrum is simple and regular
(see \re{gbmn1a}, \re{gbmn1bmass} and \re{gbmn1dmass}). 
In the second situation, there is no associated scalar tower. 
The allowed vector mass spectra	are much richer  
(see Tables \re{tablescaltw}, \re{tablescal0}, \re{tablescaltac}) and similar to the spectra found 
for the scalar and spinor towers (see the first and fourth articles of \cite{GN1}). 

It may be too naive to try to associate the towers for vector states
that we have obtained from a flat five-dimensional compactified space 
to experimentally existing particle masses
\re{physmassexp}. 
Indeed, as is well-known, the mass scale given by the parameter $1/R$, 
which should be of the order of Tev for a flat space, 
suffers from severe defects \re{physmass}. 
Randall and Sundrum \cite{RS1} elegantly cured this problem 
by the consideration of warped spaces with 
an exponential behavior of the metric in the extra dimension. 
We intend to address this issue in a forthcoming article. 

Regarding the photon, it is believed that it
has no associated massless scalar partner,  
as an axion like particle \cite{WW} has up to now escaped
observation. If it is then
a low lying state of a vector tower, 
boundary conditions of the Generic Case 2 type must apply. 
Moreover, the zero mass of the photon would require very precise relations (fine tuning!)
between the free parameters, including the bulk mass. 

\vskip 1.5 cm

\noindent{\bf{Acknowledgment:}}
The authors would like to thank Professor David Fairlie for a discussion 
about general solutions of the equations of motion and Professor Nicolas Boulanger for 
discussions in relation with particles of higher spins.

\newpage

\begin{appendix}

\section{Discussion of the joined 
boundary conditions for $B_{\{n\}}^{[V]}$ and $B_{\{n\}}^{[S]}$ 
resulting from the gauge conditions \label{joinbound}} 

The a priori independent
allowed sets of boundary conditions 
for $B_{\{n\}}^{[S]}$ and $B_{\{n\}}^{[V]}$
are given in Table \re{tablebounds}. 
The gauge conditions \re{eqgauge} imply that  
these functions 
and hence their respective boundary conditions
are related (see \re{connectgen1}-\re{connect2}).

\subsection{Generic Case 1 \label{gencase}}

In the Generic Case 1
\re{connectgen1}, the functions $B_{\{n\}}^{[V]}$ and $B_{\{n\}}^{[S]}$ are related by
\be
\partial_s B_{\{n\}}^{[S]}=B_{\{n\}}^{[V]}.
\label{apconnectbs}
\ee
This implies the following compatibility conditions

\begin{description}

\item{$\bullet$ Set A1 for $B_{\{n\}}^{[V]}$ (see Table \re{tablebounds}).}

If the boundary conditions for $B_{\{n\}}^{[V]}$ are of the form A1 
\bea
B_{\{n\}}^{[V]}(2\pi R)&=&\alpha_1^{[V]} B_{\{n\}}^{[V]}(0)
    +\alpha_2^{[V]}\partial_sB_{\{n\}}^{[V]}(0) 
             \nonumber          \\
\partial_s B_{\{n\}}^{[V]}(2\pi R)&=&\alpha_3^{[V]} B_{\{n\}}^{[V]}(0)
                    +\alpha_4^{[V]}\partial_sB_{\{n\}}^{[V]}(0) 
              \nonumber         \\
&&\alpha_1^{[V]}\alpha_4^{[V]}-\alpha_2^{[V]}\alpha_3^{[V]}=1,
\label{apcasea1}
\eea
those deduced naively from \re{apconnectbs} for $B_{\{n\}}^{[S]}$, taking into account
\re{eqbs}, give also boundary conditions of the form A1 for $B_{\{n\}}^{[S]}$ with
\bea
\alpha_1^{[S]}&=&\alpha_4^{[V]}
     \nonumber\\
\alpha_2^{[S]}&=&-\frac{\alpha_3^{[V]}}{v_{\{n\}}^2}    
     \nonumber\\
\alpha_3^{[S]}&=&-{v_{\{n\}}^2}\, {\alpha_2^{[V]}}
     \nonumber\\
\alpha_4^{[S]}&=&\alpha_1^{[V]}.
\label{aprela1}
\eea
Since the boundary parameters must be independent of $\{n\}$, and taking into account that
$\alpha_1^{[S]}\alpha_4^{[S]}{-}\alpha_2^{[S]}\alpha_3^{[S]}$ must be equal to 1,
the values of all the parameters are given in terms of one parameter only, say $\alpha$
\bea
&&\alpha_2^{[S]}=\alpha_3^{[S]}=\alpha_2^{[V]}=\alpha_3^{[V]}=0
     \nonumber\\
&&\alpha_1^{[V]}=\alpha\ \ ,\ \  
{\alpha_4^{[V]}}=\frac{1}{\alpha}\ \ ,\ \  
\alpha_1^{[S]}=\frac{1}{\alpha}\ \ ,\ \  
\alpha_4^{[S]}=\alpha
\label{aprela1bis}
\eea
leading to the set G1a in \re{caseG1}.

\item{$\bullet$ Set A2 for $B_{\{n\}}^{[V]}$ (see  Table \re{tablebounds}).}

Applying the same reasoning as for the set A1, one finds that the only allowed case A2 for
$B_{\{n\}}^{[V]}$ is with $\rho_1^{[V]}=0$ and $\rho_2^{[V]}=0$. 
It implies the case
A5 for $B_{\{n\}}^{[S]}$.
The set G1c in \re{caseG1} follows.

\item{$\bullet$ Set A3 for $B_{\{n\}}^{[V]}$ (see Table \re{tablebounds}).}

Applying the same reasoning as for the set A1, one finds that the only allowed case A3 for
$B_{\{n\}}^{[V]}$ is with ${\kappa^{[V]}}=0$. 
It implies the case
A4 for $B_{\{n\}}^{[S]}$ with $\zeta^{[S]}=0$. The set G1d in \re{caseG1} follows.

\item{$\bullet$ Set A4 for $B_{\{n\}}^{[V]}$ (see Table \re{tablebounds}).}

Applying the same reasoning as for the set A1, one finds that the only allowed case A4 for
$B_{\{n\}}^{[V]}$ is with $\zeta^{[V]}= 0$. 
It implies the case A3 for 
$B_{\{n\}}^{[S]}$ with $\kappa^{[S]}= 0$.  The set G1e in \re{caseG1} follows.

\item{$\bullet$ Case A5 for $B_{\{n\}}^{[V]}$ (see Table \re{tablebounds})}.

Applying the same reasoning as for the set A1, one finds that the only allowed case for 
$B_{\{n\}}^{[S]}$ arising from the case A5 for
$B_{\{n\}}^{[V]}$ is the case A2 with $\rho_1^{[S]}=\rho_2^{[S]}=0$.
The set G1b in \re{caseG1} follows. 

\item
{$\bullet$ The results for the Generic Case 1} are summarized in \re{caseG1} in Sect.\re{flatBCfinal}.

\end{description}


\subsection{Generic Case 2 \label{gencase2}}

The Generic Case 2, 
corresponding  to $B_{\{n\}}^{[V]}\neq 0$ and $B_{\{n\}}^{[S]}=0$ \re{connectgen2}, 
is compatible with the 
boundary conditions of all the sets of Table \re{tablebounds} for $[P]=[V]$.

\subsection{Special Cases 1 and 2\label{specase1}}

The Special Cases 1 and 2, 
corresponding to the function $B_{\{n\}}^{[S]}=1$ \re{connect1} and respectively
to the function $B_{\{n\}}^{[V]}\neq 0$ or $B_{\{n\}}^{[V]}= 0$, 
are compatible with any allowed boundary conditions for $B_{\{n\}}^{[V]}$ and with
two subsets only of boundary conditions for $B_{\{n\}}^{[S]}$ (originating from the Sets A1 and A2 of Table \re{tablebounds}) 
\bea
&{\rm{Boundary\ Conditions\ Sa}} \left\{ 
                  \begin{array}{rcl}
           B_{\{n\}}^{[S]}(2\pi R)&=&B_{\{n\}}^{[S]}(0)+\alpha_2^{[S]} \partial_s B_{\{n\}}^{[S]}(0)        \cr
           \partial_s B_{\{n\}}^{[S]}(2\pi R)&=& \partial_s B_{\{n\}}^{[S]}(0)        \cr
               \end{array}   
               \right. &
        \label{apcaseS1a} \\        
&\hspace{-3.75 cm}{\rm{Boundary\ Conditions\ Sb}} \left\{
               \begin{array}{rcl}
           \partial_s B_{\{n\}}^{[S]}(0)     &=&0        \cr
           \partial_s B_{\{n\}}^{[S]}(2\pi R)&=&0        \cr
               \end{array}
               \right. .&    
        \label{apcaseS1b}         
\eea

\end{appendix}

\newpage

\begin{table}
\caption{Table of allowed boundary conditions for $B_{\{n\}}^{[P]}$
with $[P]=[S]$ or $[P]=[V]$, 
for all $\{n\}$, from \re{BCrel1}or \re{BCrel2}. 
}
{\label{tablebounds}}
\vspace{1 cm}
\hspace{2 cm}
\begin{tabular}{|c|l|}
\hline
  Set  &\quad \quad Boundary Conditions for $B_{\{n\}}^{[P]}$ 
      \\ \hline\hline
    A1              &$B_{\{n\}}^{[P]}(2\pi R)\ \ \ =\alpha_1^{[P]}\,B_{\{n\}}^{[P]}(0)
                                   +\alpha_2^{[P]}\,\partial_sB_{\{n\}}^{[P]}(0) $
                       \\
                    & $\partial_sB_{\{n\}}^{[P]}(2\pi R)=\alpha_3^{[P]}\,B_{\{n\}}^{[P]}(0)
                    +\alpha_4^{[P]}\,\partial_sB_{\{n\}}^{[P]}(0) $
                       \\
                    & \quad\quad \quad\quad$\alpha_1^{[P]}\alpha_4^{[P]}-\alpha_2^{[P]}\alpha_3^{[P]}=1$
       \\ \hline
    A2              & $\partial_sB_{\{n\}}^{[P]}(0)\quad\ =\rho_1^{[P]}\,B_{\{n\}}^{[P]}(0) $
                       \\
                    & $\partial_sB_{\{n\}}^{[P]}(2\pi R)=\rho_2^{[P]}\,B_{\{n\}}^{[P]}(2\pi R) $
       \\ \hline
    A3              & $B_{\{n\}}^{[P]}(0)\quad \quad\ =0 $
                       \\
                    & $\partial_sB_{\{n\}}^{[P]}(2\pi R)=\kappa^{[P]}\, B_{\{n\}}^{[P]}(2\pi R) $
       \\ \hline
    A4              &  $B_{\{n\}}^{[P]}(2\pi R)=0 $
                       \\    
                    & $\partial_sB_{\{n\}}^{[P]}(0)\ =\zeta^{[P]}\, B_{\{n\}}^{[P]}(0) $  
       \\ \hline
    A5              & $B_{\{n\}}^{[P]}(0)\quad\ =0 $
                       \\
                    & $B_{\{n\}}^{[P]}(2\pi R)=0 $
      \\ \hline
\end{tabular}
\end{table}


\begin{table}
\caption{Table of Kaluza-Klein vector towers $v_{\{n\}}^2>0$ for the Generic Case~2 
($B_{\{n\}}^{[V]}(s)=\sigma_{\{n\}}^{[V]} \sin(v_{\{n\}} s)+\tau_{\{n\}}^{[V]} \cos(v_{\{n\}} s)$, $B_{\{n\}}^{[S]}(s)=0$).
The upper index $[V]$ has been omitted for the parameters $\sigma_{\{n\}}^{[V]},\tau_{\{n\}}^{[V]},\alpha_1^{[V]},\dots$\ .}
{\label{tablescaltw}}
\vspace{1 cm}
\hspace{-1cm}
\tiny{
\begin{tabular}{|l|l|l|}
\hline
\multicolumn{3}{|c|}
      {Generic Case 2 ($B_{\{n\}}^{[S]}=0$, see \re{connectgen2})}  \\
\multicolumn{3}{|c|}
       {${\phantom{\biggl[\biggr.}}$Boundary Conditions and Towers for Real Vector Fields with $v_{\{n\}} \geq 0$}
        \\ \hline \hline
    Case&Boundary condition &Tower equations
      \\ \hline\hline
   gc2-A1-tw & $B_{\{n\}}^{[V]}(2\pi R)\quad=\alpha_1B_{\{n\}}^{[V]}(0)
   +\alpha_2\partial_sB_{\{n\}}^{[V]}(0)$ &
                    $v_{\{n\}}\!\left(\alpha_1{+}\alpha_4\right)\cos\left(2\pi v_{\{n\}} R\right)
                    +\left(\alpha_2 v_{\{n\}}^2{-}\alpha_3\right)\sin\left(2\pi v_{\{n\}} R\right)
                    -2v_{\{n\}}=0$
     \\
              &$\partial_sB_{\{n\}}^{[V]}(2\pi R)=\alpha_3B_{\{n\}}^{[V]}(0)
              +\alpha_4\partial_s B_{\{n\}}^{[V]}(0)$
             & \quad\quad $\left(\sin\left(2\pi v_{\{n\}} R\right){-}\alpha_2v_{\{n\}}\right)\sigma_{\{n\}}
             =\left(\alpha_1{-}\cos\left(2\pi v_{\{n\}} R\right)\right)\tau_{\{n\}}    $
       \\
              &\quad\quad$\alpha_1\alpha_4-\alpha_3\alpha_2=1$ &
      \\ \hline
   gc2-A2-tw &  $\partial_s B_{\{n\}}^{[V]}(0)\quad\ =\rho_1\, B_{\{n\}}^{[V]}(0)$
                       &
                   $v_{\{n\}}\left(\rho2-\rho_1\right)\cos(2\pi v_{\{n\}} R)
                   +\left(v_{\{n\}}^2+\rho_1\rho_2\right)\sin(2\pi v_{\{n\}} R)=0$
       \\
            &  $\partial_s B_{\{n\}}^{[V]}(2\pi R)=\rho_2\,B_{\{n\}}^{[V]}(2\pi R)$&
                 \quad\quad $v_{\{n\}}\sigma_{\{n\}}=\rho_1\tau_{\{n\}}$
   \\ \hline
    gc2-A3-tw &  $B_{\{n\}}^{[V]}(0)\quad\quad\ =0$  &   $v_{\{n\}}\cos(2\pi v_{\{n\}} R)-\kappa\sin(2\pi v_{\{n\}} R)=0$
        \\
              &  $\partial_sB_{\{n\}}^{[V]}(2\pi R)=\kappa\, B_{\{n\}}^{[V]}(2\pi R)$&
              \quad\quad   $\tau_{\{n\}}=0$
      \\ \hline
    gc2-A4-tw &  $B_{\{n\}}^{[V]}(2\pi R)=0$ &   $v_{\{n\}}\cos(2\pi v_{\{n\}} R)+\zeta\sin(2\pi v_{\{n\}} R)=0$
    \\
             & $\partial_sB_{\{n\}}^{[V]}(0)\ =\zeta\, B_{\{n\}}^{[V]}(0)$ &
             \quad\quad $v_{\{n\}}\sigma_{\{n\}}=\zeta\,\tau_{\{n\}}$
     \\ \hline
    gc2-A5-tw & $B_{\{n\}}^{[V]}(0)\quad\ =0$ &   $\sin(2\pi v_{\{n\}} R)=0$
         \\
             & $B_{\{n\}}^{[V]}(2\pi R)=0$ & \quad\quad $\tau_{\{n\}}=0$
      \\ \hline
\end{tabular}
   }
\end{table} 


\begin{table}
\caption{Table of Kaluza-Klein vector states $v_{\{n\}}^2=0$ for the Generic Case~2 
($B_{\{n\}}^{[V]}(s)=A_{\{n\}}^{[V]} s+B_{\{n\}}^{[V]}$, $B_{\{n\}}^{[S]}(s)=0$).
The upper index $[V]$ has been omitted for the parameters $A_{\{n\}}^{[V]},B_{\{n\}}^{[V]},\alpha_1^{[V]},\dots$\ .}
{\label{tablescal0}}
\vspace{1 cm}
\hspace{1cm}
\scriptsize{
\begin{tabular}{|l|l|l|}
\hline
\multicolumn{3}{|c|}
      {Generic Case 2 ($B_{\{n\}}^{[S]}=0$, see \re{connectgen2})}  \\
\multicolumn{3}{|c|}
       {${\phantom{\biggl[\biggr.}}$Boundary Conditions for a Real Vector State with $v_{\{n\}} = 0$}
        \\ \hline \hline
    Case&Boundary condition &Tower equations
      \\ \hline\hline
   gc2-A1-0 & $B_{\{n\}}^{[V]}(2\pi R)\quad=\alpha_1B_{\{n\}}^{[V]}(0)
   +\alpha_2\partial_sB_{\{n\}}^{[V]}(0)$ &
                   $(\alpha_1+\alpha_4-2)-2\pi R\,\alpha_3=0$
     \\
              &$\partial_sB_{\{n\}}^{[V]}(2\pi R)=\alpha_3B_{\{n\}}^{[V]}(0)
              +\alpha_4\partial_s B_{\{n\}}^{[V]}(0)$
             & \quad\quad $(1-\alpha_4)\,A=\alpha_3\,B   $
       \\
              &\quad\quad$\alpha_1\alpha_4-\alpha_3\alpha_2=1$ &
      \\ \hline
   gc2-A2-0 &  $\partial_s B_{\{n\}}^{[V]}(0)\quad\ =\rho_1\, B_{\{n\}}^{[V]}(0)$
                       &
                   $(\rho1-\rho_2)-2\pi R\,\rho_1\rho_2=0$
       \\
            &  $\partial_s B_{\{n\}}^{[V]}(2\pi R)=\rho_2\,B_{\{n\}}^{[V]}(2\pi R)$&
                 \quad\quad $A=\rho_1\,B$
   \\ \hline
    gc2-A3-0 &  $B_{\{n\}}^{[V]}(0)\quad\quad\ =0$  &   $2\pi R\,\kappa-1=0$
        \\
              &  $\partial_sB_{\{n\}}^{[V]}(2\pi R)=\kappa\, B_{\{n\}}^{[V]}(2\pi R)$&
              \quad\quad   $B=0$
      \\ \hline
    gc2-A4-0 &  $B_{\{n\}}^{[V]}(2\pi R)=0$ &   $2\pi R\,\zeta+1=0$
    \\
             & $\partial_sB_{\{n\}}^{[V]}(0)\ =\zeta\, B_{\{n\}}^{[V]}(0)$ &
             \quad\quad $A=\zeta\,B$
     \\ \hline
    gc2-A5-0 & $B_{\{n\}}^{[V]}(0)\quad\ =0$ &  {\rm{impossible}} 
         \\
             & $B_{\{n\}}^{[V]}(2\pi R)=0$ & \quad\quad 
      \\ \hline
\end{tabular}
   }
\end{table} 


\begin{table}
\caption{Table of Kaluza-Klein vector state $v_{\{n\}}^2<0$ ($w_{\{n\}}^2=-v_{\{n\}}^2>0$) for the Generic Case 2 
($B_{\{n\}}^{[V]}(s)=\sigma_{\{n\}}^{[V]} \sinh(w_{\{n\}} s)+\tau_{\{n\}}^{[V]} \cosh(w_{\{n\}} s)$, $B_{\{n\}}^{[S]}(s)=0$).
The upper index $[V]$ has been omitted for the parameters $\sigma_{\{n\}}^{[V]},\tau_{\{n\}}^{[V]},\alpha_1^{[V]},\dots$\ .}
{\label{tablescaltac}}
\vspace{1 cm}
\hspace{-1.5cm}
\tiny{
\begin{tabular}{|l|l|l|}
\hline
\multicolumn{3}{|c|}
      {Generic Case 2 ($B_{\{n\}}^{[S]}=0$, see \re{connectgen2})}  \\
\multicolumn{3}{|c|}
       {${\phantom{\biggl[\biggr.}}$Boundary Conditions and Towers for a Real Vector State with $w^2_{\{n\}}=-v_{\{n\}}^2 > 0$}
        \\ \hline \hline
    Case&Boundary condition &Tower equations
      \\ \hline\hline
   gc2-A1-tac & $B_{\{n\}}^{[V]}(2\pi R)\quad=\alpha_1B_{\{n\}}^{[V]}(0)
   +\alpha_2\partial_sB_{\{n\}}^{[V]}(0)$ &
                    $w_{\{n\}}\!\left(\alpha_1{+}\alpha_4\right)\cosh\left(2\pi w_{\{n\}} R\right)
                    -\left(\alpha_2 w_{\{n\}}^2{+}\alpha_3\right)\sinh\left(2\pi w_{\{n\}} R\right)
                    -2w_{\{n\}}=0$
     \\
              &$\partial_sB_{\{n\}}^{[V]}(2\pi R)=\alpha_3B_{\{n\}}^{[V]}(0)
              +\alpha_4\partial_s B_{\{n\}}^{[V]}(0)$
             & \quad\quad $\left(\sinh\left(2\pi w_{\{n\}} R\right){-}\alpha_2w_{\{n\}}\right)\sigma_{\{n\}}
             =\left(\alpha_1{-}\cosh\left(2\pi w_{\{n\}} R\right)\right)\tau_{\{n\}}    $
       \\
              &\quad\quad$\alpha_1\alpha_4-\alpha_3\alpha_2=1$ &
      \\ \hline
   gc2-A2-tac &  $\partial_s B_{\{n\}}^{[V]}(0)\quad\ =\rho_1\, B_{\{n\}}^{[V]}(0)$
                       &
                   $w_{\{n\}}\left(\rho_1-\rho_2\right)\cosh(2\pi w_{\{n\}} R)
                   +\left(w_{\{n\}}^2-\rho_1\rho_2\right)\sinh(2\pi w_{\{n\}} R)=0$
       \\
            &  $\partial_s B_{\{n\}}^{[V]}(2\pi R)=\rho_2\,B_{\{n\}}^{[V]}(2\pi R)$&
                 \quad\quad $w_{\{n\}}\sigma_{\{n\}}=\rho_1\tau_{\{n\}}$
   \\ \hline
    gc2-A3-tac &  $B_{\{n\}}^{[V]}(0)\quad\quad\ =0$  &   $w_{\{n\}}\cosh(2\pi w_{\{n\}} R)-\kappa\sinh(2\pi w_{\{n\}} R)=0$
        \\
              &  $\partial_sB_{\{n\}}^{[V]}(2\pi R)=\kappa\, B_{\{n\}}^{[V]}(2\pi R)$&
              \quad\quad   $\tau_{\{n\}}=0$
      \\ \hline
    gc2-A4-tac &  $B_{\{n\}}^{[V]}(2\pi R)=0$ &   $w_{\{n\}}\cosh(2\pi w_{\{n\}} R)+\zeta\sinh(2\pi w_{\{n\}} R)=0$
    \\
             & $\partial_sB_{\{n\}}^{[V]}(0)\ =\zeta\, B_{\{n\}}^{[V]}(0)$ &
             \quad\quad $w_{\{n\}}\sigma_{\{n\}}=\zeta\,\tau_{\{n\}}$
     \\ \hline
    gc2-A5-tac & $B_{\{n\}}^{[V]}(0)\quad\ =0$ &   $\sinh(2\pi w_{\{n\}} R)=0\ \rightarrow\ ${\rm{impossible}}
         \\
             & $B_{\{n\}}^{[V]}(2\pi R)=0$ & \quad\quad $\tau_{\{n\}}=0$
      \\ \hline
\end{tabular}
   }
\end{table} 

\end{document}